\documentclass[twocolumn,aps,prl,superscriptaddress]{revtex4-1}

\usepackage{color,amsmath,amssymb,epsfig,graphicx}
\usepackage{bm}
\usepackage{mathptmx, courier, pifont}
\usepackage[scaled=0.92]{helvet}
\usepackage[T1]{fontenc}
\usepackage{textcomp}
\usepackage[colorlinks=true,linkcolor=blue,filecolor=blue,urlcolor=blue,citecolor=blue]{hyperref}

\begin{document}
\author{S.~Frauendorf}
\email{sfrauend@nd.edu}
\affiliation{Department of Physics and Astronomy, University of Notre Dame, Notre Dame, Indiana 46556,  USA}
\title{Chirality in rotating nuclei}
\date{\today}
\begin{abstract}
The article explains on a non-expert level how triaxial  rotating nuclei may attain a chiral structure.
\end{abstract}
\maketitle

\begin{figure}[h]
\center{\includegraphics[angle=-90,width=\linewidth,trim=0 0 0 0 ,clip]{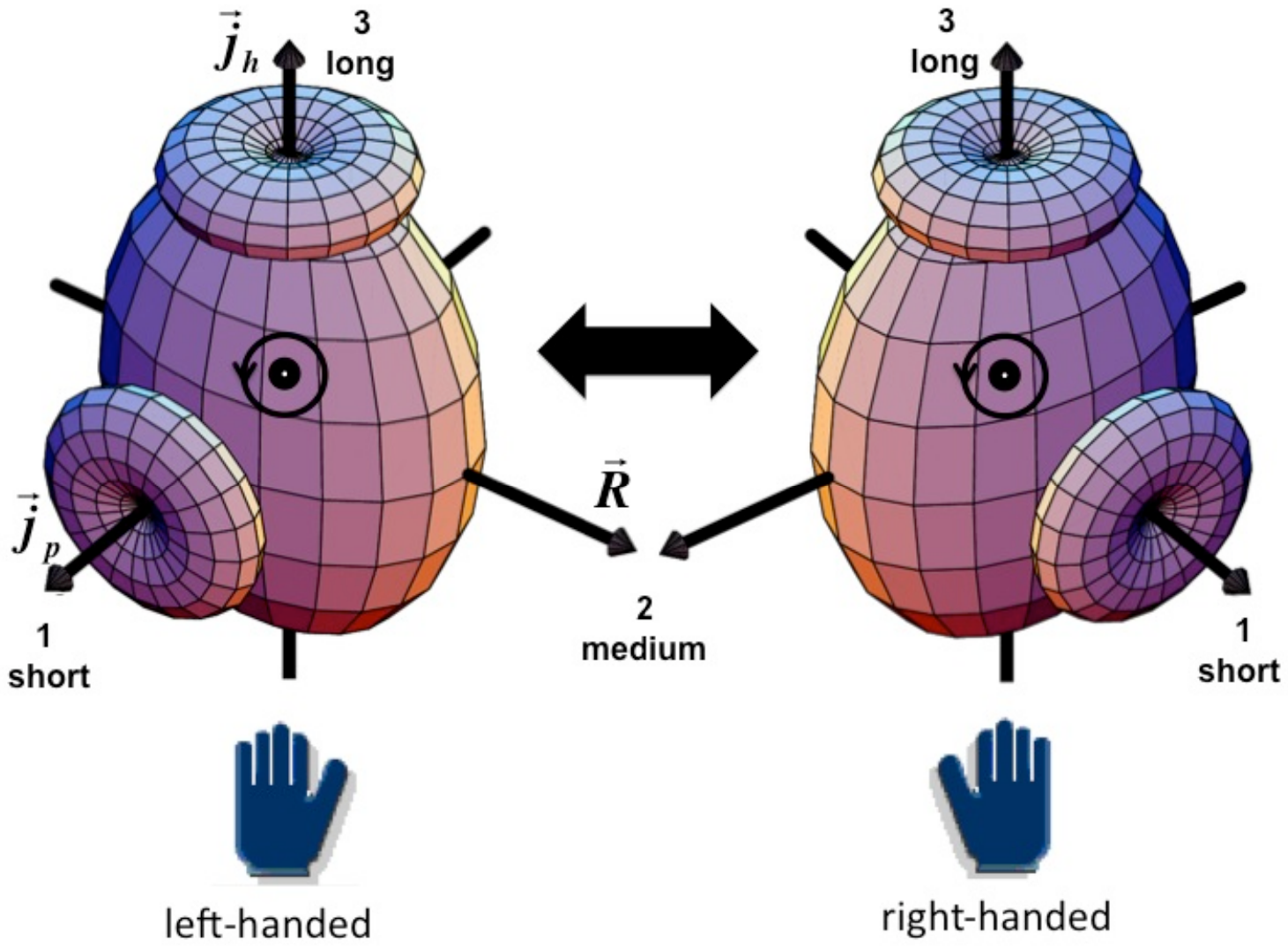}}
 \caption{\label{f:shape} A rotating triaxial nucleus with an unpaired particle carrying the angular momentum $\vec{j}_p$, an unpaired hole carrying $\vec{j}_h$ , and collective angular momentum $\vec{R}$ .  All angular momenta are displayed as arrows. The total angular momentum  $\vec{J}$, which is the
 axis of rotation,  points into the direction of sight and is depicted by the circle symbolizing the arrow tip. For purpose of illustration the orbitals are drawn outside the nucleus, whereas they are located inside.
 }
\end{figure}

{\bf Chiral molecules.}
Chirality (Greek "handiness") is a property of many complex molecules. Chiral molecules exist in two forms, one being the mirror image of the other. Like for our hands, it is impossible to make the images identical by a suitable rotation. (For simple, achiral molecules, as H$_2$O, this is possible.) The two forms are called left-handed and right-handed. They have the same binding energy, because the electromagnetic interaction, which holds the molecule together, does not change under a reflection. Other properties that are insensitive to the geometry are also the same. The different geometry is the reason why the left-handed form turns the polarization plane of transmitted light in one direction by some angle while of the right- handed form turns it in the opposite direction by the same angle. The differences between the two species may have dramatic consequences.  
\begin{figure}[h]
\center{\includegraphics[angle=-90,width=\linewidth,trim=0 0 0 0 ,clip]{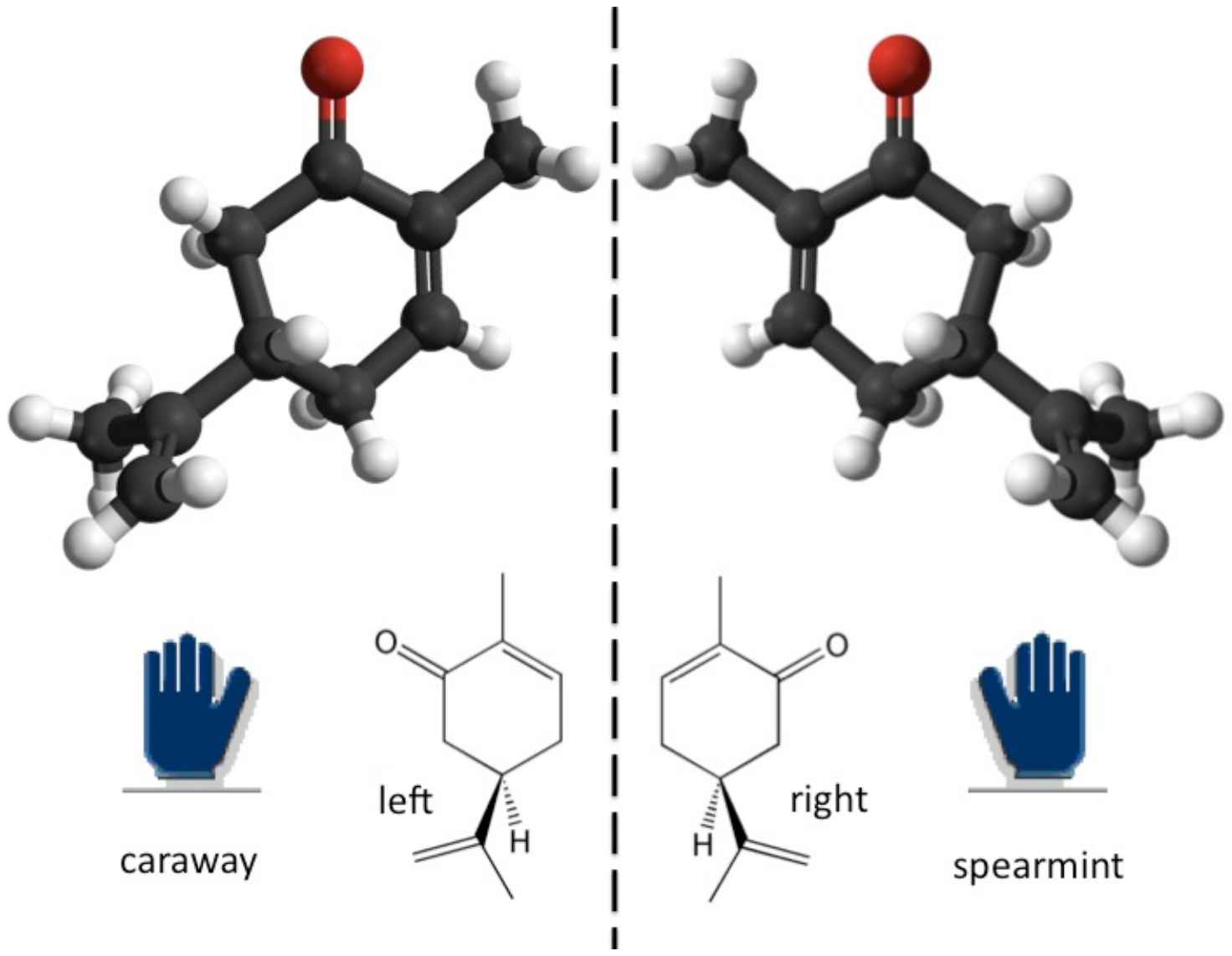}}
 \caption{\label{f:Carvon} The two chiral configurations of the carvon molecule. Black balls represent the C  atoms, white the H  atoms, and red the O  atom. 
 The left-handed molecule is transformed into the right-handed by a reflection through the mirror plane, which is   perpendicular to the figure plane
and goes through the broken line.
 }.
\end{figure}
Fig. \ref{f:Carvon}  shows the carvon molecule. Its left-handed form smells of caraway and its right-handed form of spearmint. Organisms usually synthesize only the left-handed or the right-handed species of a molecule. 
{\bf Shapes of atomic nuclei.}
Compared to molecules, nuclei are very compact. They resemble deformed droplets of  liquid, with nearly constant interior density and a thin surface. This reflects the short range of the interaction that binds the nucleons together.  Inside, the nucleons move on orbitals whose shapes are determined by the quantization of angular momentum. As an example, Fig. \ref{f:shape} shows the orbitals that carry an angular momentum of $\frac{11}{2}\hbar$.
The origin of nuclear deformation is the shape of the quantized orbitals of the nucleons, just as  the shape of the electronic orbitals leads to the complex structure of molecules.  However, there is a difference. The repulsive long-range Coulomb force between the electrons localizes the low-j orbitals (carrying $\frac{1}{2}\hbar$  or $\frac{3}{2}\hbar$) on the outside of the atoms, where they lobe out making chemical bonds with neighboring atoms. In contrast, the short-range attractive force between the nucleons pulls the nucleonic orbitals carrying $\frac{1}{2}\hbar\leq j\leq\frac{11}{2}\hbar$ as close as possible together, such that the different shapes of the orbitals nearly average out.  The resulting shapes are simple: spherical, axial symmetric, mostly football-like, sometimes pear-shaped. Some nuclei take the shape of a triaxial ellipsoid. Such simple shapes do not attain the property of chirality, which was considered as an alien concept by nuclear physicists.  

{\bf Rotation of nuclei.}
As with molecules, deformed nuclei show rotational spectra. These are sequences of states of quantized energy and angular momentum, which have the same structure, differing from each other only by the angular velocity with respect to the axis of rotation. Concerning the dynamics of rotation, molecules behave like classical bodies. The arrangement of the nuclei of the constituent atoms determines the three moments of inertia according to classical mechanics. Uniform rotation is only possible about the two principal axes with the largest and the smallest moment of inertia. In contrast, the dynamics of nuclear rotation is governed by the quantized motion of the individual nucleons, which makes nuclei react like a clock work of gyroscopes, where the gyroscopes are the nucleonic orbitals carrying a fixed angular momentum  $\vec{j}$.  In even-even nuclei (even number of protons and even number of neutrons) being in the lowest energy state, the nucleons occupy pair wise orbitals with opposite directions of 
 $\vec{j}$. Rotation is only possible if the shape is asymmetric with respect to the axis of rotation. Because of this arrangement, even-even nuclei with an axially symmetric shape only rotate about an axis perpendicular to their symmetry axis. Even-even triaxial nuclei rotate about the medium axis, which has the largest moment of inertia, because the deviation from axial symmetry is maximal (one of the perpendicular axes is the short one and the other is the long one).
 
{\bf  Tilted rotational axis.} A new pattern evolves if not all  $\vec{j}_p$ of the orbitals are paired off, which appears in odd-odd nuclei (odd number of protons and odd number neutrons). Unlike molecules, such nuclei may rotate about an axis that is tilted with respect to the axes characterizing its shape. Fig. \ref{f:shape} illustrates such a situation for a triaxial nucleus. The large shape
represents the "core" of the nucleus, which contains all the nucleons but the odd proton and has one more neutron, with all orbitals paired.  
The odd proton occupies the doughnut-like orbital, which is added to the core. The angular momentum of the orbital $\vec{j}_p$  points along the short axis of the core. This orientation corresponds to the energy minimum of the short-range attractive interaction between the orbital and the core. The presence of the odd neutron hole can be seen as the horizontal  doughnut-like orbital carved out  from in the even-even core (which has one more neutrons). The angular momentum  $\vec{j}_h$ of this hole-orbital points along the long axis of the core, because this orientation corresponds to the energy minimum of the short-range repulsive interaction between the hole-orbital and the core. (The opposite sign of the hole-core interaction is analogous 
to the opposite sign of the gravitational pull on a bubble in water, its buoyancy.)
The core behaves like an even-even triaxial nucleus, generating a component $\vec{R}$ along the medium axis. The total angular momentum $\vec{J}=\vec{j}_p+\vec{j}_h+\vec{R}$, which is the axis of rotation, is tilted respect to the three axes of the triaxial shape.  

{\bf Emergence of chirality.} The appearance of a tilted rotational axis had been known for axial nuclei, when S. Frauendorf and J. Meng \cite{FM97} first noticed that in a case like shown in Fig. \ref{f:shape}, the rotating triaxial nucleus attains chirality: Looking from the tip of the vector $\vec{J}$ onto the shape, the short, medium, long half-axes are ordered clockwise in the right-handed configuration and counterclockwise in the left-handed configuration. Although the two configurations in Fig. \ref{f:shape} look like mirror images, they are {\bf not} related by a reflection as in the case of molecules. 
A reflection does not change the angular momenta ($\vec{R}=m\vec{v}\times\vec{r}$). One has to invert the direction $\vec{R}$, which selects the half-axis, in order to convert the left-handed into the right-handed configuration. (An additional rotation brings it into the position shown in the figure.) This is achieved by time reversal, which changes only the sign of the velocities. The nucleons run "backwards" ($\vec{R}=m\vec{v}\times\vec{r}$). There are three equivalent chiral pairs generated by changing the directions of $\vec{j}_p$ and $\vec{j}_h$, which are combined with the shown pair to superposition determined by the $I$, the quantum number pf $J$.

The existence of two distinct configurations gives rise to two bands of rotational states where the states  with the same angular momentum are expected to have
the same energy and  emit 
$\gamma$-radiation in the same way. The chiral partners  have the same parity quantum number, 
which indicates that the configurations do not change the under reflections through the principal planes
of the triaxial shape. A more detailed discussion of the symmetries of rotating nuclei can be found in Ref. \cite{Frau01}.
As discussed below, the coupling between the left- and right-handed configurations perturbs the ideal pattern to some extend.

\begin{figure}[t]
\center{\includegraphics[angle=-90,width=\linewidth,trim=0 0 0 0 ,clip]{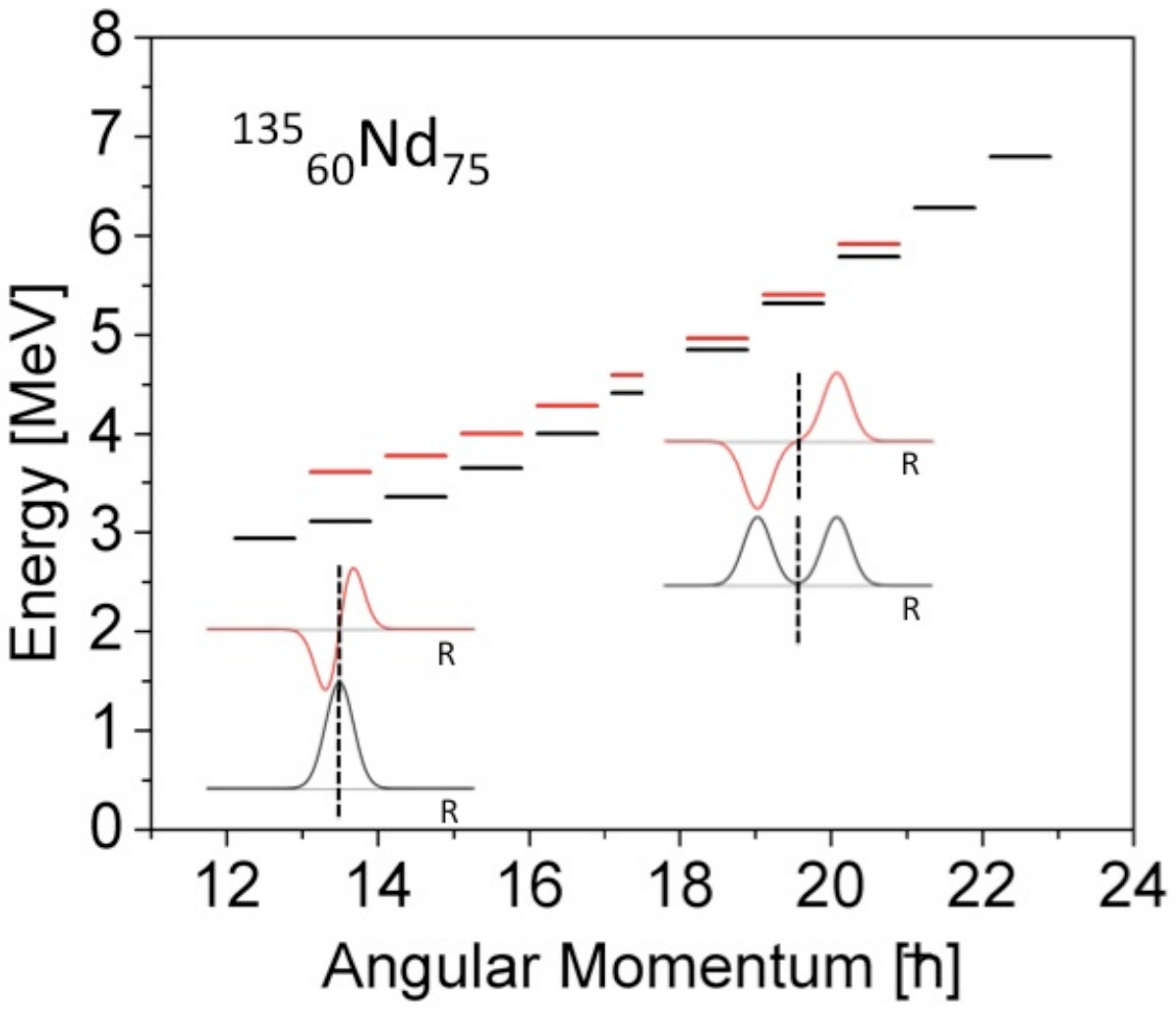}}
 \caption{\label{f:spectrum} The two energy sequences of rotational states in the nucleus $^{135}$Nd that constitute chiral pairs.  The insets illustrate how the left-handed and right-handed configurations are coupled. 
 The broken lines depict the  short-long plane perpendicular to the
 figure plane. The double arrows show $ \vec{j}_p$, $\vec{j}_h$ in the short- long plane. 
 The curves show the wave function of the collective angular momentum component along the medium axis. The square of the wave function is probability distribution of this component.
 }
\end{figure}

 Candidates for rotational sequences of such chiral doublets have first been found in odd-odd nuclei with approximately 134 nucleons and approximately 104 nucleons.  The presence of chirality was best demonstrated in $^{135}$Nd by the observation of both doublets and equal emission of  $\gamma$-radiation from the two states \cite{Zhu03,Muk07}. This nucleus, shown in Fig. \ref{f:spectrum}, has an even number of 60 protons. Chirality sets in only at sufficiently large angular momentum when a proton pair with opposite $ \vec{j}_p$ is broken by flipping the direction of  $ \vec{j}_p$ of one proton, such that both  $ \vec{j}_p$ are aligned with the short axis. Yet another type of chiral configuration was found in the even-even nucleus $^{112}$Ru \cite{Luo09}. Two neutron pairs are broken, which generate angular momentum both along the short and long axes.

{\bf Left-right coupling.} The left- and right-handed configurations are coupled. For molecules this coupling is usually so weak that once formed, 
a right-handed configuration remains such for any time span of practical concern. 
There are exceptions: CH$_3$NHF, oscillates with a frequency of 1000GHz between right and left. In this case, the left- and right-handed configurations differ only 
by the location of one electron pair, which are separated by a low barrier.

For nuclei the coupling the left-right coupling is substantial and depends on the total angular momentum.  As seen in Fig. \ref{f:shape}, the left-handed configuration 
differs from the right-handed configuration by the orientation  of the component $\vec{R}$ along the medium axis. In order to convert chirality, 
the vector $\vec{R}$ must turn through the plane that contains the long and short axes. The energy is higher when $\vec{R}$   lies in this plane than when $\vec{R}$  is
 aligned with the medium-length axis. 
 The energy difference, which increases with $R$, is the barrier that decouples the left-handed from the right-handed configuration. 
 Since $R\left(=\sqrt{J^2-j_p^2-j_h^2}\right)$ increases with increasing $J$, the left-right coupling decreases with $J$. 
 
 The coupling generates two stationary states, which are the even and odd  combinations of the left- and right-handed configurations. They have
 somewhat different energies, where the difference increases with the coupling strength. Fig. \ref{f:spectrum}  shows the sequence of such chiral pairs. Their energy distance reflects the decrease 
  of the left-right coupling with increasing $J$. For $J=\frac{27}{2}\hbar$, the nucleus oscillates rapidly between left and right (chiral vibration).  
  The wave functions are the ones of the ground and first excited states of a harmonic oscillator. 
  For $J=\frac{39}{2}\hbar$, the chirality is strongest. The left- and right-handed components of the wave function are well separated and the energies of the two bands 
  approach each other. The nucleus turns three times around before it changes from left to right (static chirality).  
 
 For background information see CHIRALITY; NUCLEAR STRUCTURE; STRONG NUCLEAR INTERACTION; RIGID-BODY DYNAMICS in the McGraw-Hill Encyclopedia of Science \& Technology
 and the review article \cite{Frau01}.


\begin{thebibliography}{99}

\bibitem{FM97} S. Frauendorf, J. Meng, Tilted rotation of triaxial nuclei, Nucl. Phys. A 617:131, 1997.

\bibitem{Frau01} S. Frauendorf, Spontaneous symmetry breaking in rotating nuclei, Rev. Mod. Phys. 73:463, 2001.

\bibitem{Zhu03} S. Zhu {\it et al.}, A composite chiral pair of rotational bands in the odd-A nucleus Nd-135, Phys. Rev. Lett. 91:132501, 2003.

\bibitem{Muk07} S. Mukhopadhyay {\it et al.}, From chiral vibration to static chirality in Nd-135, Phys. Rev. Lett. 99: 172501, 2007. 

\bibitem{Luo09} X. Y. Luo {\it et al.}, Evolution of chirality from gamma Soft Ru-108 to triaxial Ru-110, Ru-112, Phys. Lett. B 670:307, 2009.


\end{thebibliography}
\end{document}